\documentstyle[aps]{revtex}
\newcommand{\ds}{\displaystyle}
\newcommand{\dsf}{\ds\frac}

\newcommand{\beq}{\begin{equation}}
\newcommand{\eeq}{\end{equation}}

\large

\begin{document}
\large
\begin{center}
\Large\bf
Quasi-Stationary Temperature Profile and Magnetic Flux Jumps
in Hard Superconductors
\vskip 0.1cm
{\normalsize\bf N.A.\,Taylanov}\\
\vskip 0.1cm
{\large\em Theoretical Physics Department,\\
Scientific Research Institute for Applied Physics,\\
National University of Uzbekistan,\\
Tashkent, 700174, Uzbekistan\\
E-mail: taylanov@iaph.tkt.uz}
\end{center}
\begin{center}
{\bf Abstract}
\end{center}
\begin{center}
\mbox{\parbox{14cm}{\small

In the present paper, the temperature distribution in the critical state
of hard superconductors is investigated in the quasi-stationary approximation.
It is shown that the temperature profile can be essentially inhomogeneous
in the sample, which affects the conditions of initiation of a magnetic
flux jumps.
}}
\end{center}
\vskip 3mm

While dealing with instabilities of the critical state in hard
superconductors, the character of the temperature distribution $T(x,t)$
and that of the electromagnetic field $\vec E(x,t)$ are of substantial
practical interest [1]. This derives from the fact that thermal and magnetic
destractions of the critical state caused by Joule self-heating are
defined by the initial temperature and electromagnetic field distributions.
Hence, the form of the temperature profile may noticeably influence the
criteria of critical-state stability with respect to jumps in the magnetic
flux in a superconductor. Earlier (cf., e.g., [2]), in dealing with
this problem, it was usually assumed that the spatial distribution of
temperature and field were either homogeneous or slightly inhomogeneous.
However, in reality, physical parameters of superconductors may be
inhomogeneous along the sample as well as in its cross-sectional plane.
Such inhomogeneities can appear due to different physical reasons.
First, the vortex structure pinning can be inhomogeneous due to the
existence of weak bonds in the superconductor. Second, inhomogenety
of the properties may be caused by their dependence on the magnetic
field $H$. Indeed, the field $H$ influences many physical quantities,
such as the critical current density $j_c$, the differential conductivity
$\sigma_d$, and the heat conductivity $k$.

In the present paper, the temperature distribution in the critical state
of hard superconductors is investigated in the quasi-stationary approximation.
It is shown that the temperature profile can be essentially inhomogeneous,
which affects the conditions of initiation of a magnetic flux jumps.

The evolution of thermal ($T$) and electromagnetic
($\vec E, \vec H$) perturbations in superconductors is described by a
nonlinear heat conduction equation [3],
\beq
\nu\dsf{dT}{dt}=\nabla [\kappa\nabla T]+\vec j\vec E,
\eeq
a system of Maxwell's equations,
\beq
rot\vec E=-\dsf{1}{c}\dsf{d\vec H}{dt},
\eeq
\beq
rot{\vec H}=\dsf{4\pi}{c}\vec j
\eeq
and a critical-state equation
\beq
\vec j=\vec j_{c}(T,\vec H)+\vec j_{r}(\vec E).
\eeq
Here $\nu=\nu(T)$ is the specific heat, $\kappa=\kappa(T)$ is the thermal
conductivity respectively; $\vec j_c$ is the critical current density and
$\vec j_r$ is the active current density.            

We use the Bean-London critical state model to
describe the $j_{c}(T,H)$ dependence, according to which
$j_{c}(T)=j_0[1-a(T-T_{0})]$ [4], where the parameter $a$ characterizes
thermally activated weakening of Abrikosov vortex pinning on crystal
lattice defects, $j_{0}$ is the equilibrium current density, and
$T_0$ is the temperature of the superconductor.

The $j_r(E)$ dependence in the region of sufficiently strong electric
fields $(E\ge E_f$; where $E_f$ is the limit of the linear region of the
current-voltage characteristic of the sample [2]) can be approximated
by a piecewise-linear function $j_r\approx\sigma_f E$, where
$\sigma_f=\dsf{\eta c^2}{H\Phi_0}\approx \dsf{\sigma_n H_{c_2}}{H}$ is the
effective conductivity in the flux flow regime and $\eta$ is the viscous
coefficient, $\Phi_0=\dsf{\pi h c}{2e}$ is the magnetic flux quantum,
$\sigma_n$ is the conductivity in the normal state, $H_{c_2}$ is the upper
critical magnetic field. In the region of the weak fields $(E\le E_f)$, the function $j_r (E)$ is
nonlinear. This nonlinearity is associated with thermally activated
creep of the magnetic flux [5].

Let us consider a superconducting sample placed into an external magnetic
field $\vec H=(0, 0, H_{e})$ increasing at a constant rate
$\dsf{d\vec H}{dt}=\dot H$=const. According to the Maxwell equation (2),
a vortex electric field $\vec E=(0, E_e, 0)$ is present. Here $H_e$ is
the magnitude of the external magnetic field and $E_e$ is the magnitude
of the back-ground electric field. In accordance with the concept of
the critical state, the current density and the electric field must be
parallel: $\vec E\parallel \vec j$;.
       The thermal and electromagnetic boundary conditions for the
Eqs. (1)-(4) have the form
\beq
\begin{array}{ll}
\left.\kappa \dsf{dT}{dx}\right|_{x=0}+w_0[T(0)-T_0]=0\,,\qquad &
T(L)=T_{0}\,,\\
\quad \\
\left.\dsf{dE}{dx}\right|_{x=0}=0\,, &   E(L)=0\,,
\end{array}
\eeq

For the plane geometry (Fig.) and the boundary conditions
$H(0)=H_e,\quad H(L)=0$, the magnetic field distribution is
$H(x)=H_e(L-x)$,
where $L=\dsf{cH_e}{4\pi j_c}$ is the depth of magnetic flux penetration
into the sample and $w_0$ is the coefficient of heat transfer to the
cooler at the equilibrium temperature $T_0$.

The condition of applicability of Eqs. (1)-(4) to the description of
the dynamics of evolution of thermomagnetic perturbations are discussed
at length in [2].

In the quasi-stationary approximation, terms with time derivatives can
be neglected in  Eqs. (1)-(4). This means that the heat transfer from
the sample surface compensates the energy dissipation arising in the
viscous flow of magnetic flux in the medium with an effective conductivity
$\sigma_f$.
In this approximation, the solution to Eq. (2) has the form

\beq
E=\dsf{\dot H}{c}(L-x).
\eeq

Upon substituting this expression into Eq. (1) we get an inhomogeneous
equation for the temperature distribution  $T(x,t)$,
\beq
\dsf{d^2\Theta}{d\rho^2}-\rho\Theta=f(\rho).
\eeq
Here we introduced the following dimensionless variables
$$
f(\rho)=-[1+r\omega\rho]\dsf{j_0}{aT_0},\\
\qquad
\Theta=\dsf{T-T_0}{T_0},\\
\qquad
\rho=\dsf{L-x}{r},\\
$$
and the dimensionless parameters
$\omega =\dsf{\sigma_f \dot H}{cj_0}$,
and, $r=\left(\dsf{c\kappa}{a \dot H L^2} \right)^{1/3}$,
where $r$ characterizes the spatial scale of the temperature profile
inhomogeneity in the sample.
Solutions to Eq. (7) are Airy functions, which can be expressed
through Bessel functions of the order 1/3 [6]

\beq
\Theta (\rho) = C_1\rho^{1/2}K_{1/3}\left(\dsf{2}{3} \rho^{3/2}\right)
+ C_2\rho^{1/2}I_{1/3}\left(\dsf{2}{3} \rho^{3/2}\right)+\Theta_0(\rho),
\eeq
$$
\Theta_0(\rho) = \rho^{1/2}K_{1/3}\left(\dsf{2}{3}\rho^{3/2}\right)\int_{0}^{\rho}
[1+r\omega \rho_1]\rho_{1}^{3/2}I_{1/3}
\left(\dsf{2}{3}\rho_{1}^{3/2}\right)d\rho_1 -
$$
$$
\rho^{1/2}I_{1/3}\left(\dsf{2}{3}\rho^{3/2}\right)\int_{0}^{\rho}
[1+r\omega \rho_1]\rho_{1}^{3/2}K_{1/3}
\left(\dsf{2}{3}\rho_{1}^{3/2}\right)d\rho_1,
$$
where $C_1$ and $C_2$ are integration constants, which are determined by
the boundary conditions to be
$$C_1=0,\quad C_2=\dsf{-w_0L\Theta(0)+\left.\kappa\dsf{d\Theta}{d\rho}
\right|_{\rho=\dsf{L}{r}}}
{\left.\left[w_0\left(\dsf{L}{r}\right)^{1/2}I_{1/3}
\left(\dsf{2}{3}\rho^{-3/2}\right)-
2\dsf{d}{d\rho}
\left(\rho^{1/2}I_{1/3}\left(\dsf{2}{3}\rho^{3/2}\right)\right)
\right]\right|_{\rho=\dsf{L}{r}}}
$$
From the Maxwell equation (2), the temperature inhomogeneity parameter can
be expressed in the form
\beq
\alpha=\dsf{r}{L}=\left[\dsf{4\pi\nu j_0}{aH_{e}^{2}}
\dsf{H_e}{\dot Ht_\kappa}\right]^{1/3}
\eeq
        It is evident that $\alpha\sim 1$ near the threshold for a flux jump,
when $\dsf{aH_{e}^{2}}{4\pi\nu j_0}\sim 1$, even under the quasi-stationary
heating condition $\dsf{\dot H t_{\kappa}}{H_e}<<1$; where
$t_{\kappa}=\dsf{\nu L^2}{\kappa}$ is the
characteristic time of the heat conduction problem.

       Let us estimate the maximum heating temperature $\Theta_m$
in the isothermal case $w=\dsf{\kappa}{L}\ge 1$.
The solution to Eq. (7) can be represented in the form
\beq
\Theta(x)=\Theta_m-\rho_0\dsf{(x-x_m)^2}{2},
\eeq
near the point at which the temperature is a maximum, $x=x_m$  (Fig.).

With solution (10) being approximated near the point $x_m=\dsf{L}{2}$
with the help of the thermal boundary conditions, the coefficient
$\rho_0$ can be easily determined to be $\left(\dsf{8}{L^2}\right)\Theta_m$
and the temperature can be written as
\beq
\Theta(x)=\Theta_m\left[1-\dsf{4}{L^2}\left(x-\dsf{L}{2}\right)^2\right],
\eeq

Substituting this solution into Eq.(7), the superconductor maximum
heating temperature due to magnetic flux jumps can be estimated as
\beq
\Theta_m=\dsf{\left[j_0+\dsf{\sigma_f \dot H}{c}(L-x_m)\right]\dsf{\dot H}{c \kappa T_0}
(L-x_m)}{\dsf{\gamma}{L^2}-\dsf{a\dot H}{c\kappa}(L-x_m)}.
\eeq
For a typical situation when
$\dsf{\gamma}{L^2}<<\dsf{a\dot H}{c\kappa}(L-x_m)$
the estimation for $\Theta_m$ is
\beq
\Theta_m\approx\left[j_0+\dsf{\sigma_f \dot H}{c}(L-x_m)\right]\dsf{\dot HL^2}
{c\kappa T_0}(L-x_m).
\eeq
Here, the parameter $\gamma\sim 1$ (for a parabolic temperature profile
$\gamma\sim 8$). It is easy to verify that for typical values of
$j_0=10^6 A/cm^2$,
$\dot H=10^4$ Gs/sek, and $L=0,01$ cm the heating is sufficiently low:
$\Theta_m<<1$.
In the case of poor sample cooling, $w=1-10 erg/(cm^2 s K)$, the $\Theta_m$ is
\beq
\Theta_m=\dsf{\dot Hj_0L^2}{cw_0T_0}\approx 0,5;
\eeq

i.e., the heating temperature can be as high as
$\delta T_m=T_0\Theta_m\sim 2 K$.
One can see that in the case of poor sample cooling, the heating can be
rather noticeable and influence the conditions of the thermomagnetic
instability of the critical state in the superconductor.

\newpage
\centerline{\large\bf FIGURE}

Fig. The distrubation of the temperature profile $\Theta (x)$.

\newpage
\centerline{\large \bf NIZAM A.TAYLANOV, G.B.Berdiyorov }
\begin{tabbing}
{\large \bf Address:} \\
Theoretical Physics Department and\\
Institute of Applied Physics,\\
National University of Uzbekistan,\\
Vuzgorodok, 700174, Tashkent, Uzbekistan\\
Telephone:(9-98712),461-573, 460-867.\\
fax: (9-98712) 463-262,(9-98712) 461-540,(9-9871) 144-77-28\\
e-mail: taylanov@iaph.tkt.uz \\
\end{tabbing}

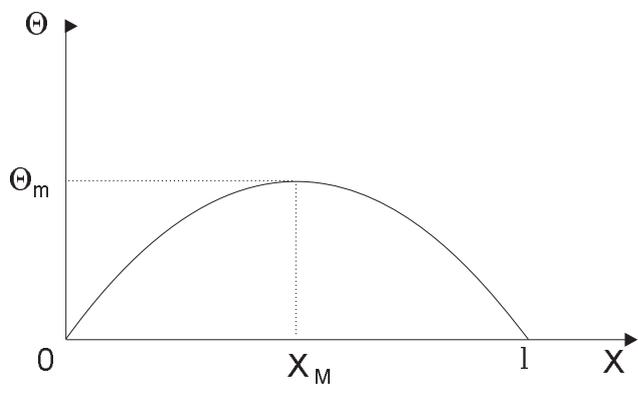
\end{document}